# Far-UVC Field Emission Device at 226 nm and its Sub-Nanometer thick GaN/AlN Quantum Well Anode

D. L. Boiko[1], P. Demolon[2], J.-F. Carlin[2], E. Eriksson[3], A. Hoogerwerf[1], K. Bach Gravesen[4], A. Brimnes Gardner[4,5], M.-A. Dubois[1], Peter Tønning[4], E. Zanchetta Ulsig[4,5], M. Marin[4], J. Tingsborg[3], N. Volet[4,5]

[1]Centre Suisse d'Electronique et de Microtechnique SA (CSEM), CH-2002 Neuchâtel, Switzerland
[2]Ecole polytechnique fédérale de Lausanne (EPFL), Lausanne, Switzerland
[3]PureFize Technologies AB, Virdings Allé 32B, 754 50 Uppsala, Sweden
[4]UVL A/S, Søren Frichs Vej 50, 8230 Åbyhøj, Denmark
[5]Aarhus University, Finlandsgade 22, 8200 Aarhus N, Denmark

We report on unique features of ultra-thin GaN/AlN quantum wells and demonstrate a new field emission device for human-safe disinfection with an external quantum efficiency of 18% and superior reliability, as compared to far-UVC LEDs. We investigate the behavior of a Shockley-Read-Hall recombination via Al vacancies – oxygen complexes in AlN barriers as a function of GaN well thickness and come to the conclusion that direct relaxation of confined states via mid-gap traps is prohibited and the main mechanism limiting the quantum efficiency (QE) is the carrier escape followed by nonradiative SRH recombination in the AlN barriers. As a consequence, the QE is highly dependent on the defect density, QW width and temperature which allowed us to reach a descent emission at 226 nm wavelength although at wall plug efficiency being low, at 0.02%. We build a complete Light-Current-Voltage model and find that present device performance is limited by the self-heating and by the Child's space charge effect, both limitations can be easily addressed with some additional engineering efforts.

Far-UVC light in the spectral range of 200-230 nm effectively kills bacteria and deactivates viruses while being not harmful to health, enabling persistent disinfection of air and surfaces in the presence of humans.[1,2] Far UVC is promising for many other applications such as stimulation of wound healing, sterilization, photocatalysis, and photochemistry etc. provided light sources exist at an affordable price. So far, only excimer lamps were a viable solution in this narrow UV range. Those light sources are commercially available but suffer from EMC compatibility issues and fundamental limitations on size-reduction potential. Far-UVC AlGaN light emitting diodes (LEDs) with wavelengths below 250 nm exhibit exponential collapse of external quantum efficiency (EQE) well below 1%.[3] Only recently, LEDs with AlGaN/AlGaN quantum wells (QWs) emitting at 226 nm were demonstrated with mW output level,[4] 0.28% EQE and 0.18% wall plug efficiency (WPE),[5] but the device lifetime is limited to ~10 h.[5]

In this work, we choose a different approach and realized field emission devices (FED) with anodes consisting of a stack of 90 monolayer-width GaN/AlN QWs. This approach allowed us to avoid all potential reliability issues linked with p-doped AlN and contact in Far-UVC LEDs.[6] Previous studies have demonstrated a battery powered Far-UVC FED with powder hexagonal Boron Nitride (hBN) anode with emission at 225nm, WPE of 0.03% at a practical 6 kV bias and the expected device lifetime of several thousands of hours.[7] The spectrum, however, showed a pronounced deep trap emission in the range of 250-350 nm, which is harmful for humans. In addition, high purity monocrystal hBN[8] entails significant cost, limiting its viability as an anode material. Another interesting demonstration of a FED at 226 nm was achieved utilizing $Zn_2SiO_4$ film anodes,[9] which however also exhibited parasitic emission at 208 nm and 244 nm spectral bands. At the same time, far UVC FEDs with AlGaN/AlGaN QWs or rare earth ions doped film anodes such as $Nd^{3+}$:$LuF_3$ thin films, demonstrated emission just above (at 240 nm[10]) or just below (at 180 nm[11]) the human-safe spectral range. Interestingly, wide bandgap hBN was also attempted to resolve the p-doping issue in AlGaN Deep UVC LEDs,[12] but as of today, we are not aware of any longlasting Far UVC AlN LED or AlN FED solutions[13] below 240nm wavelength. Difficulties are caused by exponential drop of EQE below 250 nm.[3]

In this letter we report on performance and validation in a disinfection test of FED utilizing shallow states in monolayer-width multiple GaN/AlN QW stack emitting at 226 nm. We build a full LIV performance model and discuss present limitations and possible future improvements.

Anodes with various QW widths were grown by metalorganic vapor phase epitaxy (MOVPE) on AlN templates deposited on top of c-plane oriented sapphire substrates. A typical structure comprises as much as 90 GaN monolayerwidth quantum wells (QWs) with AlN barriers of 3 nm [Fig. FIG.1(a)]. The entire structure is topped with a 10 nm thick AlN cap layer. The use of a AlN template with a dislocation density ~$10^{10}$ cm$^{-2}$ enables the emission in the range 216-260 nm by varying the QW width between 0.5 to 2 monolayers (ML) [Fig. FIG.1(b)]. For comparison, the band edge emission from AlN template is centered at 210 nm.

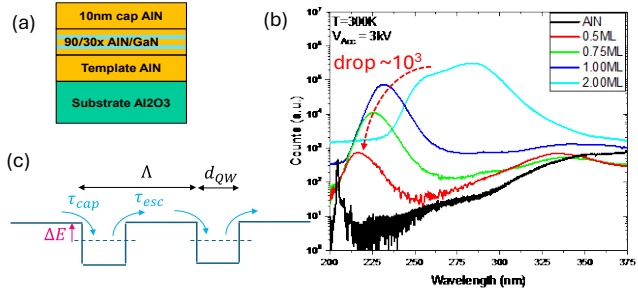

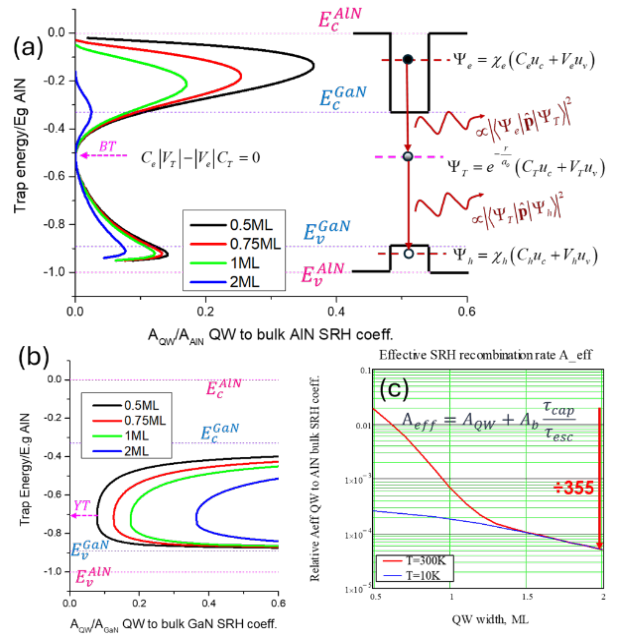

FIG.1. GaN/AlN shallow monolayer QWs: (a) epitaxial layer composition of fabricated anodes (b) cathodoluminescence of test structures with 30 shallow QWs of different monolayer thickness (color curves) grown on MOVPE AlN template (black curve). The far-UVC emission blue shift with narrowing QW width is accompanied by the exponential decay of the intensity and surfacing of the emission from the deep traps in AlN barriers (at photon energies of ~3 eV due to Shockley-Read-Hall recombination via Al vacancies – oxygen complexes). (c) Schematic representation of carrier capture and escape processes to and from QWs used to model the effect of the SRH recombination via deep traps in AlN barriers.

The energy transfer rate from the impinging high energy electron beam (EB) to nonequilibrium carriers and the scattering depth are modelled using analytical methods established at the early times of EB pumped semiconductor lasers[14] and also confirmed in Monte Carlo simulations of electron trajectory in solids (CASINO).[15] The overall active layer thickness of 0.3 μm in our GaN/AlN QW anodes is optimal for pumping by electron beam (EB) with energies of up to 8 keV considering an onset of the cathodoluminescence (CL) emission drop (at a fixed current). For comparison, a monocrystalline hBN would require ~1 μm of film growth because boron nuclei are much lighter than aluminum, resulting in a longer electron scattering length.

Monolayer GaN/AlN QW physics is very different from the famous case of infinite barrier height QWs. The energies of the confined states are just slightly below the barriers, so the tails of the wave functions extend into the barriers for tens of atomic periods. We modify the textbook example of shallow QW[16] by incorporating a perturbation due to the build-in piezoelectric field of hexagonal lattice. The second-order energy correction term [the quantum confined Stark effect (QCSE) energy shift] is a constant, explaining why the peak emission wavelength in Fig 1.(b) may not approach the AlN bandedge at 210 nm with reducing QW width.

For a 40 nm wavelength change in Fig. FIG.1(b), CL intensity drops by three orders of magnitude. We attribute the exponential drop of the quantum efficiency to the Shockley–Read–Hall (SRH) recombination. [At the carrier densities reported in this study, the Auger recombination can be ignored.]

Point defect complexes of Al vacancies and oxygen impurities are optically active leading to the blue luminescence band in bulk AlN.[17] Indeed, the black curve in Fig. FIG.1(b) for MOVPE grown AlN template with near band-gap CL at 210 nm attests for a strong SRH recombination via deep centers near the mid gap with blue luminescence band (BT) at ~400nm recombination. [A similar SRH recombination mechanism via Ga vacancies and C-related defects is responsible for the yellow luminescence band (YT) in bulk GaN.[18,19] For all studied QWs, the wavefunction envelopes largely extend into AlN barriers[16] and overlap strongly with the deep traps (DT) in the barriers. This feature cannot explain the $10^3$ efficiency drop in Fig. FIG.1(b). On the other hand, for confined states in our QWs with emitted photon energies by $\Delta E$=0.3 to 1.3 eV below the AlN gap, the capture to escape time ratio $\tau_{cap}/\tau_{esc} \propto \exp(-\Delta E/2kT)\Lambda/d_{QW}$ [20] varies by many orders of magnitude at room temperature (RT) [notation of Fig. FIG.1(c)]. These bring us to the rational that (i) direct SRH recombination of confined states via DT is largely suppressed and (ii) the carrier escape to the barriers ([Fig. FIG.1(c)]) with subsequent SRH of bulk electron and hole states via DT is the main efficiency drop mechanism. Indeed, the overall effective SRH recombination rate of confined carriers reads[21]

$$A_{eff} = A_{QW} + \frac{\tau_{cap}}{\tau_{esc}} A_b \ , \qquad (1)$$

where $A_{QW}$ and $A_b$ are, respectively, the recombination rate coefficients for the confined states in QW and for the bulk states in the barriers. Literature reports on SRH rates in AlN films and AlGaN QWs with emission below 260 nm are very sparce and contradictory. The SRH recombination coefficients differ by 4 orders of magnitude across different reports. from $A_{eff}$=1.3·$10^6$ s$^{-1}$ in Al$_{0.7}$Ga$_{0.3}$N/AlN QWs (estimated from data in Ref.[22]) to the highest reported rate for AlN films of 1·$10^{10}$ s$^{-1}$ in Ref.[23] We use the last value for $A_b$ of the barrier states in Eq.(1).

Fig. 2. Relative SRH recombination rates of confined states vs bulk $A_{QW}/A_b$ (horizontal axes) as a function of deep trap energies in AlN (a) and GaN (b) [vertical axes]. The energy is normalized on AlN bandgap. The labels BT and YT indicate, respectively, blue luminescence traps at -3 eV in AlN and yellow luminescence traps at -2.2 eV in GaN. (c): net recombination rate $A_{eff}$ [Eq.(1)] as a function of monolayer QW width at room and cryogenic temperatures.

For $A_{QW}$ of the confined states, we elaborate a simple model (more details can be found in Ref.[24]) that accounts for (i) the envelope wavefunctions overlap of the confined states[16] and hydrogenic-like DT point defects,[25,26] and (ii) band mixing of the periodic Bloch functions parts in a simplistic 2x2 $\mathbf{k} \cdot \mathbf{p}$ Hamiltonian formalism.[26] Following the classical SRH picture of defect-mediated two-step recombination process[27] we consider a matrix elements of $\hat{\mathbf{p}}$ (electric dipole transition) between confined states $\Psi_{e,h} = \chi_{e,h}(C_{e,h}u_{c0} + V_{e,h}u_{v0})$ and hydrogenic trap state $\Psi_T = e^{-r/a_0}(C_T u_{c0} + V_T u_{v0})$ [see the schematics in Fig. 2(b)], e.g.,

$$\langle \Psi_e | \hat{\mathbf{p}} | \Psi_T \rangle = \langle \chi_e | e^{-r/a_0} \rangle \left( C_e^* V_T + V_e^* C_T \right) K_p,$$   (2)

where the matric element $K_p = \langle u_{c0} | \hat{\mathbf{p}} | u_{v0} \rangle$ is taken between the conduction and valance band states at the band edge and the band mixing coefficients $C$ and $V$ for a state of some particular energy are defined from the $\mathbf{k} \cdot \mathbf{p}$ Hamiltonian. For the states in the bands above or below the gap, the coefficients $C$ and $V$ are real numbers, while in the gap, they are phase-shifted by $\pi/2$ so that if $C$ is a real number, $V$ is an imaginary one.

For the DT in AlN barriers we use periodic Bloch functions of AlN. Respectively, both confined and trap states exhibit a phase shift between the band mixing terms so as $C_e^* V_T + V_e^* C_T = i(C_e|V_T| - |V_e|C_T)$ in Eq.(2). This term (as well as the overall $A_{QW}$) vanishes for the trap energies near the mid gap, when the band mixing coefficients are of nearly equal amplitudes [see Fig. 2 (a)]. This occurs at -3 eV trap energy corresponding to BT. Strong suppression of SRH recombination of confined states via blue luminescence traps in AlN is predicted by the model for all considered QWs [In Fig. 2 (a), the SRH rate for the confined states is normalized to the rate in the bulk material $A_b = 1 \cdot 10^{10}$ s$^{-1}$].

Another contribution to $A_{QW}$ in Eq.(1) may be expected from yellow luminescence traps. Here, YT in GaN wells are analyzed using periodic Bloch functions of GaN. For the trap states, the band mixing coefficients C and V in Eq.(2) are complex numbers, but for the confined QW states (above or below the GaN gap), these are real numbers. Therefore, while the SRH recombination via YT is reduced for the confined states of the QW relative to bulk GaN ($1 \cdot 10^6$ s$^{-1}$), $A_{QW}$ never vanishes [Fig.2 (b)].

The resulting $A_{eff}$ is plotted in Fig.2 (c). At cryogenic temperatures (blue curve), the carrier escape is quenched [second term in (1)] while the remaining SRH recombination $A_{QW}$ of the confined states is much weaker than $A_b$ in AlN barriers. However, at RT (red curve) and up to 1.2 ML QW width, the carrier escape and recombination in AlN barriers largely dominate and strongly depend on the energy deficit [the second term in (1) with $\Delta E < 0.6$ eV]. For wider QWs ($\Delta E > 0.6$ eV), the direct SRH recombination $A_{QW}$ of the confined states prevails, showing weak variations with further increasing QW width $d_{QW}$. In overall, $A_{eff}$ changes by a factor of 355 between 0.5 and 2 ML QWs. Taking into account that radiative recombination coefficient scales as $B \propto d_{QW}$, one may easily see that this model retrieves 3 or-ders of the efficiency drop (see also Fig.4 (a) below), in reasonable agreement with the experimental measurements in Fig. 1 (b). In our devices utilizing 0.75 ML QWs, the efficiency is thus limited by the carrier escape followed by SRH recombination in AlN barriers.

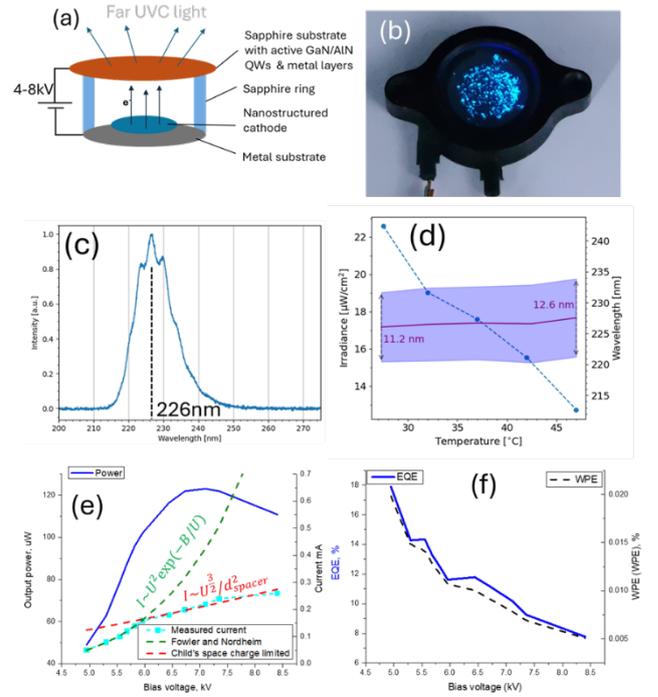

Fig. 3 (1). GaN/AlN MQW based FED: (a) Device structure and driving schematics. (b) Photographic image of operating FED. Blue emission from deep traps in AlN barriers can be seen. (c) Optical emission spectrum peaked at 226nm. (d) Thermal sensitivity of the irradiance and emitted FWHM spectral band. Measured light - current- voltage curves (e) as well as external quantum efficiency and wall plug efficiency versus bias voltage (f). The performance of the reported devices is limited by the self-heating (thermal resistance 15 K/W) and Child's space charge effect at a bias above 6 kV preventing acceleration of electrons.

The anodes with QWs of 0.75ML width were integrated into tablet-shaped FEDs with nanostructured cold-emission cathodes [Fig.3(a)]. The number of QW was increased from 30 [in Fig 1.(b)] to 90 to allow for efficient pumping with EB energies up to 8 keV without significant DT emission from AlN template. Visual inspection of operating FEDs reveals a weak BT emission from the circular emission area (12mm in diameter) [Fig.3 (b)]. The main strong far UVC emission is at 226 nm wavelength [Fig.3 (c)] and there is no harmful emission in the rest of UV spectrum. Due to carrier escape from shallow QWs, the device emission is highly sensitive to temperature showing an intensity drop by a factor of 1.8 for a 20°C change [Fig.3 (d)]. Importantly, the central emission wavelength remains below the cut off 230 nm while FWHM slightly increases with temperature.

The device demonstrates a record EQE of 18% (the number of far-UVC photons per 100 electrons emitted from cathode) thanks to multiple scattering of high energy electrons in the anode material [Fig. 3(f)]. The power reaches up to 120 μW at 226 nm, albeit with a modest wall plug

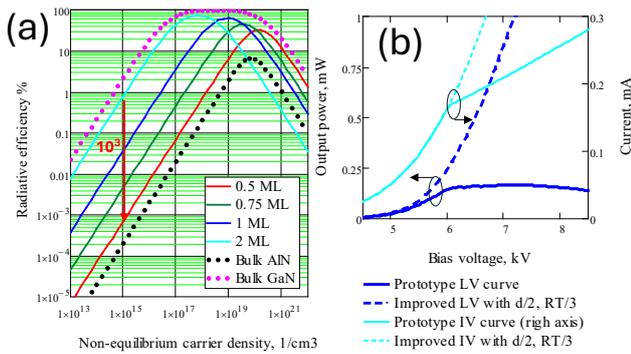

Fig. 4. (a): Radiative efficiency[3] vs carrier density in the QWs of different width in comparison with the bulk AlN and GaN. (b): Modelled LIV curves in tested prototype [solid curves, compare to measurements shown in Fig. 3 (e)] and predicted for device with 3 times lower thermal resistance $R_T$ and 2 times thinner anode-cathode spacer.

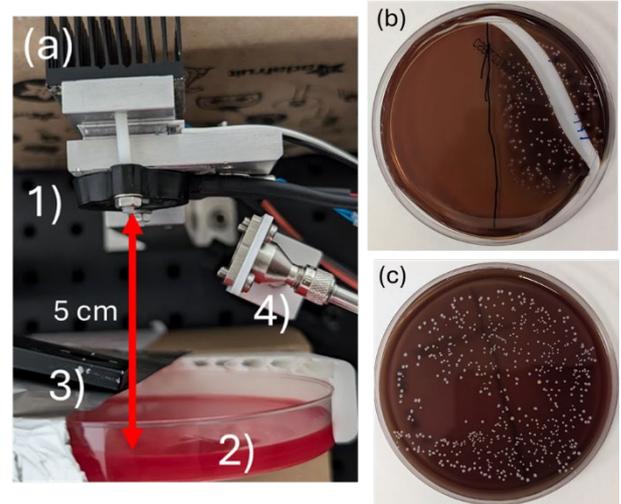

Fig. 5. Validation of UVC disinfection features on general hand bacteria (staphylococcus epidermidis): (a) Photographic image of the test setup showing AlN FED (1), half masked Agar plate with bacteria (2), power meter (3) and fiber spectrometer input (4). The exposure time is 7h, incubation time is 32h: (a) left hand side of Agar plate was subjected to illumination and right-hand side was masked; (b) control plate with no illumination.

efficiency of 0.02%. The output power as a function of bias voltage exhibits rollover.

To better understand these limitations, we developed a comprehensive light-current-voltage model [Fig 4.(b), solid curves]. It shows that the performance is primarily constrained by the self-heating (15 K/W thermal resistance) and the Child's space charge effect[28] at a bias above 6 kV. Both factors can be mitigated through targeted engineering improvements. The model predicts the output power of 1 mW can be reached at 7 kV bias if the thermal resistance is reduced to 5 K/W and the anode-cathode spacer is shrunk by a factor of 2. Such improvement is possible because of the radiative efficiency[3] scaling linearly with carriers (Fig.4 (a)). Preliminary biological test confirms effective disinfection of general hand bacteria (staphylococcus epidermidis). The far-UVC FED was 5 cm from half-covered Agar plate with bacteria and delivered, over 7 h, a total dose of 9.36 mJ/cm² (Fig 5 (a)). After incubation for 32 h, no colony growth is visible in the illuminated half plate (Fig. 5(b)), contrary to the covered half-plate region and a control sam-ple (Fig 5 (c)). Most importantly, in these tests we noticed that the lifetime of our FEDs emitting at 226 nm is well above the tens of hours scale (further studies are needed to define the device lifetime).

In summary, we demonstrate a far-UVC FED for human-safe disinfection applications that operates at 226 nm. The device utilizes ultra-thin GaN/AlN QWs grown by MOVPE and exhibits nearly 20% EQE. Further performance improvement can be easily achieved via physics package design optimization.

Authors are deeply grateful to Nicolas Grandjean for valuable discussions and MOVPE facilities. This study was carried out with the financial support from the EUROSTARS project SAFEUVC (Project ID E!531).